\documentclass[prd,twocolumn,showpacs,preprintnumbers,amsmath,amssymb,superscriptaddress,floatfix,nofootinbib]{revtex4}
\usepackage{txfonts}
\usepackage{graphicx}
\usepackage{amsmath}
\usepackage{amsfonts}
\usepackage{amssymb}
\usepackage{color}
\usepackage{multirow}
\usepackage[colorlinks, citecolor=blue,anchorcolor=red,menucolor=red,linkcolor=red,filecolor=red,runcolor=red,urlcolor=blue,frenchlinks=red]{hyperref}

\usepackage{subfigure}

\begin{document}

\title{The $a_0(980)$ in the single Cabibbo-suppressed process $\Lambda_c \to \pi^0\eta p$}

\author{Xue-Chao Feng}
\affiliation{College of Physics and Electronic Engineering, Zhengzhou 
University of Light Industry, Zhengzhou 450002, China}

\author{Le-Le Wei}
\affiliation{Institute of Particle Physics and Key Laboratory of Quark and Lepton Physics (MOE), Central China Normal University, Wuhan, Hubei 430079, China}

\author{Man-Yu Duan\footnote{Corresponding author }}
\email{duanmy@seu.edu.cn}
\affiliation{School of Physics, Southeast University, Nanjing 210094, China}

\author{En Wang\footnote{Corresponding author }}
\email{wangen@zzu.edu.cn}
\affiliation{School of Physics and Microelectronics, Zhengzhou University, Zhengzhou, Henan 450001, China}
\affiliation{Guangxi Key Laboratory of Nuclear Physics and Nuclear Technology, Guangxi Normal University, Guilin 541004, China}

\author{De-Min Li}
\affiliation{School of Physics and Microelectronics, Zhengzhou University, Zhengzhou, Henan 450001, China}

\begin{abstract}
In this work, we have investigated the Cabibbo-suppressed process $\Lambda_c \to \pi^0\eta p$, by taking into account the intermediate scalar state $a_0(980)$, which could be dynamically generated from the $S$-wave pseudoscalar-pseudoscalar interaction within the chiral unitary approach. We have calculated the $\pi^0\eta$ invariant mass distribution, and found that there is a significant structure associated to the $a_0(980)$. We have also roughly estimated the branching fraction $\mathcal{B}(\Lambda_c \to \pi^0\eta p) \sim 10^{-4}$.  We encourage our experimental colleagues to measure the process $\Lambda_c \to \pi^0\eta p$ for searching for the state $a_0(980)$ in this reaction.

\end{abstract}



\maketitle

\section{Introduction}
\label{sec:intro}
Investigating the dynamic structure of resonances is one of the important directions in hadron physics, and the decay mechanism of the lightest charm baryon $\Lambda_c$ is of great significance to understand the interplay of the weak and strong interaction in the charm region~\cite{Cheng:2015iom,Ebert:1983ih,
Lu:2016ogy,Geng:2018upx}. In the last decades, lots of experimental information have been accumulated~\cite{BESIII:2018mes,LHCb:2018mes,Belle:2018obs,Belle:2017sea,BESIII:2019mes}, and there exist many theoretical studies about $\Lambda_c$ decay~\cite{Chao-Qiang Geng:2019sin,Hai-Yang Cheng:2018sin,Xiao-Hai Liu:2019vis,Jung Keun Ahn:2019hyp}.
 
Recently, the BESIII Collaboration has reported the $K^+K^-$ invariant mass distribution of the process $\Lambda_c\to p K^+K^-$, which shows an enhancement structure near the $K^+K^-$ threshold~\cite{BESIII:2018mes}.  In Ref.~\cite{Wang:2020}, we have analyzed the experimental measurement of the $\Lambda_c\to p K^+K^-$, and concluded that the enhancement structure near the threshold is mainly due to the resonance $f_0(980)$, and $a_0(980)$ provides a small contribution. 
Since both $a_0(980)$ and $f_0(980)$ couple to the $K\bar{K}$, the mechanisms of their productions are crucial for us to deeply understand about the $K\bar{K}$ enhancement in the process $\Lambda_c\to p K^+K^-$ and other processes. For instance, the LHCb Collaboration has argued that the $a_0(980)$ plays a more important role than the $f_0(980)$ in decay $\bar{B}^0\to J/\psi K^+K^-$, and reported the branching fraction $\mathcal{B}(\bar{B}^0\to J/\psi a_0(980), a_0(980)\to K^+K^-)=(4.70\pm 3.31\pm 0.27)\times 10^{-7}$~\cite{Aaij:2013mtm}, however, both the $f_0(980)$ and $a_0(980)$ resonances are expected to contribute to the $K^+K^-$ distribution, as discussed in Ref.~\cite{Liang:2015qva}. 

Taking into account the uncertainties of the experimental measurements and the undefined theoretical parameters, it is still difficult to extract the relative weight of the $a_0(980)$ and $f_0(980)$ in the processes involving $K\bar{K}$ final states, which is necessary for us to understand the production mechanisms of $a_0(980)$ and $f_0(980)$ in the $K^+ K^-$ channel. Since the $a_0(980)$ mainly decays to $\pi\eta$ channel, we propose to investigate the process $\Lambda_c\to \pi^0\eta p$, which should be useful to understand the production mechanism of the $a_0(980)$. 
The reaction of $\gamma p \to p \pi^0\eta$ has been measured by Crystal Barrel at the electron stretcher accelerator ELSA, where the $a_0(980)$ is clearly identified~\cite{Gutz:2014wit}. Up to our knowledge, the  process $\Lambda_c\to \pi^0\eta p$ has not yet been investigated theoretically and experimentally. 

%
%

For  the $a_0(980)$ nature, there are many theoretical explanations, such as tetraquark, molecular state~\cite{Weinstein:1990gu,Baru:2003qq,Hooft:2008we} (see the review `Scalar mesons below 2~GeV' of Particle Data Group (PDG)~\cite{PDG2022}).  The internal structure of the $a_0(980)$ is essential to establish the decay mechanism of the process $\Lambda_c\to \pi^0 \eta p$, and is also crucial  to understand the spectrum of the scalar mesons with isospin $I=1$~\cite{Wang:2017pxm}.  Among those different explanations, the molecular nature of the $a_0(980)$, dynamically generated from the $S$-wave pseudoscalar-pseudoscalar interaction within the chiral unitary approach~\cite{Oller:1997chi,Nieves:1999bet,Cabrera:2005eva,Guo:2006dyn}, has been widely studied in literatures, such as $\bar{B}^0\rightarrow J/\psi\pi^{0}\eta$~\cite{Liang:2015qva},   $\eta_{c} \rightarrow \eta\pi^{+}\pi^{-}$~\cite{Debastiani:2017}, $D^0\rightarrow \bar{K}^0\pi^{0}\eta$~\cite{Xie:2014tma},
$D^+\to \pi^+\pi^0\eta$~\cite{Duan:2020vye}, $D_s^+\to K^+K^-\pi^+$~\cite{Wang:2021naf,Wang:2021nxz,Zhu:2022wzk}, $D_s^+\to  K_S^0K^+\pi^0$~\cite{Zhu:2022guw}, $D^+_s\to \pi^+\pi^0\eta$~\cite{Molina:2019udw,Hsiao:2019ait}, $J/\psi\to \gamma \pi^0\eta$~\cite{Xiao:2019lrj},  $\Lambda_c \to p K^+K^-$~\cite{Wang:2020}, and $\Lambda_c\to \Lambda \eta\pi$~\cite{Wang:2022nac}.


In this work, we will investigate the process $\Lambda_c\to \pi^0\eta p$, considering the $a_0(980)$ dynamically generated from the $S$-wave pseudoscalar-pseudoscalar interaction in the chiral unitary approach. By predicting the $\pi^0\eta$ invariant mass distribution and the branching fraction of this reaction, we would like to provide a motivation for experimentalists to measure this process. The experimental information of this process could be useful to learn about the nature of $a_0(980)$, and also constrain the parameters that is crucial to understand the mechanism of the $a_0(980)$ and $f_0(980)$ production in the $K^+K^-$ system.  

The paper is organized as follows. In Sect.~\ref{sec:form}, we introduce our model for  the process $\Lambda_c \to  \pi^0\eta p$, which can be divided into three steps, weak process,  hadronization, and the final state interaction. The results and discussions are given in Sect.~\ref{sec:results}, a short summary is given in the last section.

\section{Formalism}  \label{sec:form}

\begin{figure*}[tbhp]
\begin{center}
\includegraphics[scale=0.4]{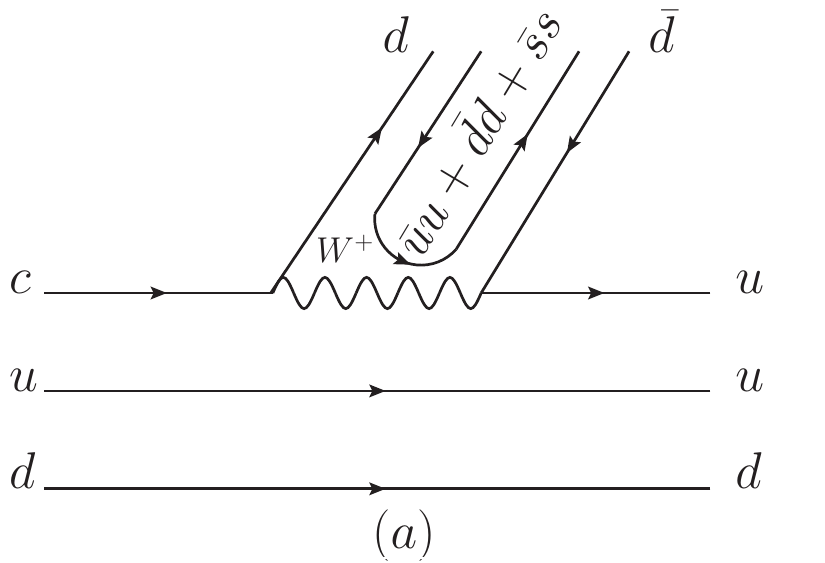}
\includegraphics[scale=0.4]{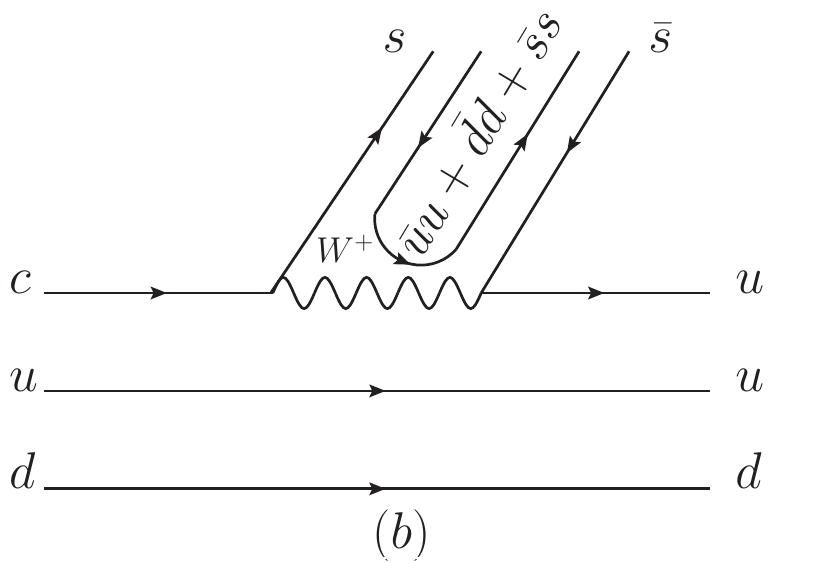}
\includegraphics[scale=0.4]{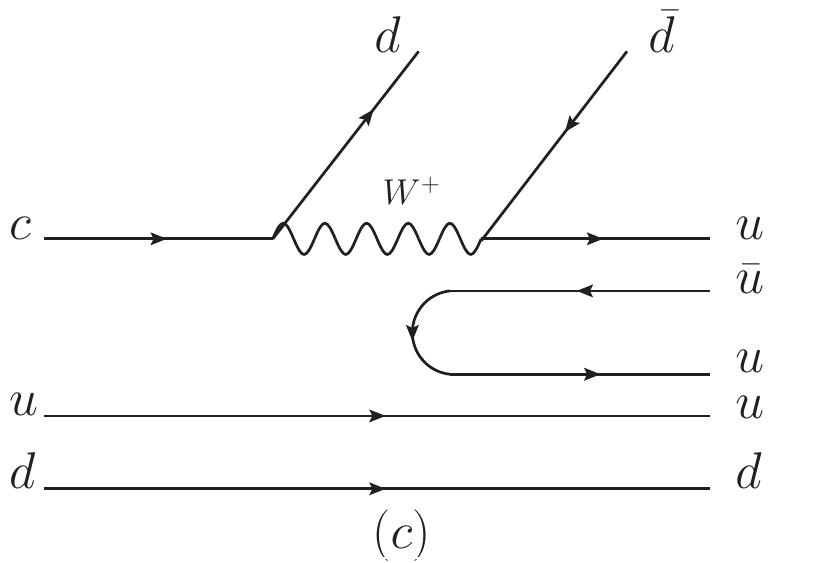}
\includegraphics[scale=0.4]{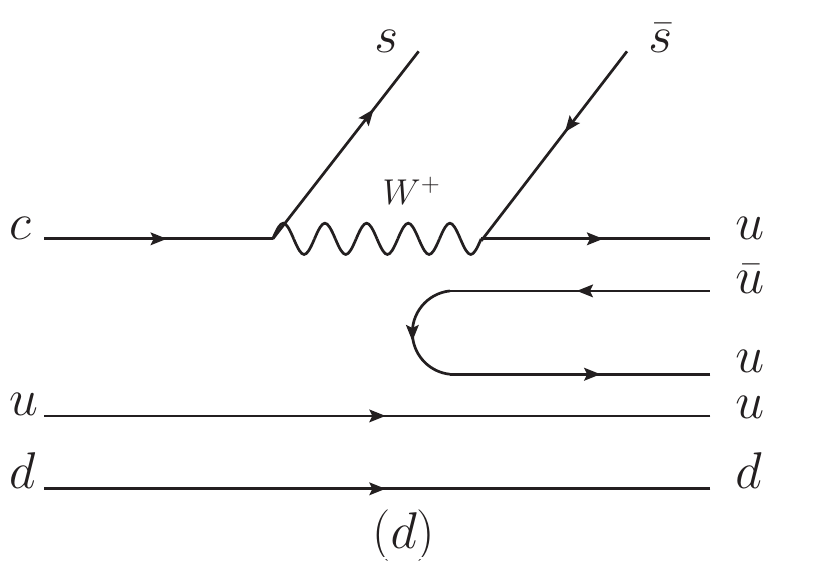}
\includegraphics[scale=0.4]{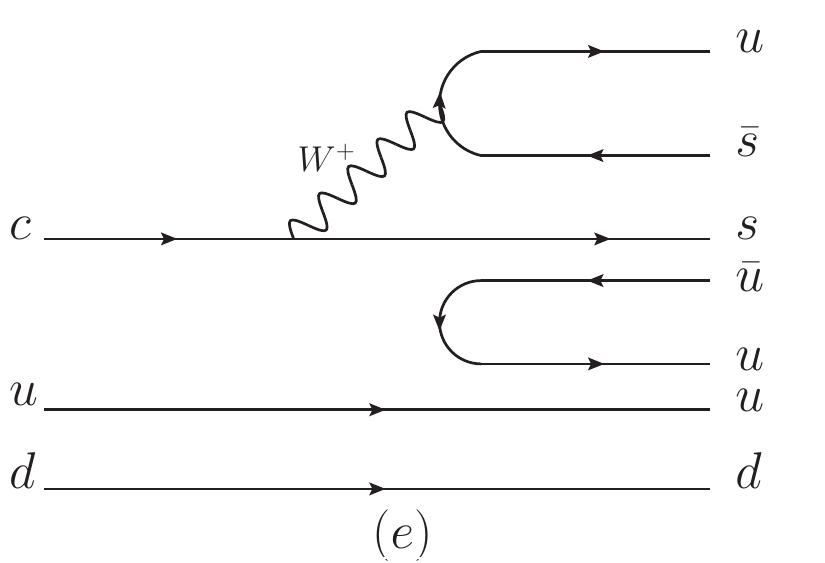}
\end{center}
\caption{The microscopic quark level diagrams for the process
$\Lambda_c\rightarrow \pi^0\eta p$, (a) the $W^+$ internal emission ($c\to dW^+\to d\bar{d}u$) with $\bar{q}q$ produced from the vacuum between $d$ and $\bar{d}$, (b) the $W^+$ internal emission ($c\to sW^+\to s\bar{s}u$) with $\bar{q}q$ produced from the vacuum between $s$ and $\bar{s}$, 
(c) the $W^+$ internal emission ($c\to dW^+\to d\bar{d}u$) with $\bar{u}u$ produced from the vacuum between $u$ and $ud$, (d) the $W^+$ internal emission  ($c\to sW^+\to s\bar{s}u$) with $\bar{u}u$ produced from the vacuum between $u$ and $ud$, and (e) the $W^+$ external emission ($c\to sW^+\to s\bar{s}u$) with  $\bar{u}u$ produced from the vacuum between $s$ and $ud$.}
\label{Fig:quarkthreebody}
\end{figure*}

In analogy to Ref.~\cite{Wang:2020}, the single Cabibbo-suppressed decay of $\Lambda_c\rightarrow \pi^0\eta p$ can proceed via three steps: the weak process, the hadronization, and the final state interaction. Firstly, the $c$ quark of the initial $\Lambda_c$ weakly decays into a $W^{+}$ boson and a $d$ (or $s$) quark, then $W^{+}$ boson turns into $u\bar{d}$ (or $u\bar{s}$) pair. In order to give rise to the final states of the process $\Lambda_c\rightarrow \pi^0\eta p$, the $ud$ pair of the initial $\Lambda_c$ and the $ud\bar{d}$ (or $us\bar{s}$) quarks from the weak decay must be hadronized with $\bar{q}q$ ($\equiv \bar{u}u+\bar{d}d+\bar{s}s$) created from vacuum with $J^{PC}=0^{++}$, which can be classified as the $W^+$ internal emission and $W^+$ external emission, respectively depicted in Figs.~\ref{Fig:quarkthreebody}(a)-(d) and Fig.~\ref{Fig:quarkthreebody}(e).

For the $W^+$ internal emission mechanisms of Figs.~\ref{Fig:quarkthreebody}(a) and \ref{Fig:quarkthreebody}(b), the $d\bar{d}$ or $s\bar{s}$ quarks from the weak decay are hadronized with $\bar{q}q$ ($\equiv \bar{u}u+\bar{d}d+\bar{s}s$) created from vacuum. It should be pointed out that, in Fig.~\ref{Fig:quarkthreebody}(b), the $s\bar{s}$ pair with isospin $I=0$, together with the created $\bar{q}q$ pair, can not be hadronized into the $\pi^0\eta$ system with isospin $I=1$, which implies that the mechanism of Fig.~\ref{Fig:quarkthreebody}(b) has no contribution to the process $\Lambda_c \to \pi^0\eta p$. For Fig.~\ref{Fig:quarkthreebody}(a), after hadronization we have the flavor wave function of the final state from the $\Lambda_c$ weak decay,
\begin{eqnarray}
\Lambda_c &=& \frac{1}{\sqrt{2}}c(ud-du) \nonumber \\
&\Rightarrow & V_{cd}V_{ud} d\left(\bar{u}u+\bar{d}d+\bar{s}s\right)\bar{d} u\frac{1}{\sqrt{2}}\left(ud-du\right) \nonumber \\
&=&  V_{cd}V_{ud} \sum_i \left(M_{2i}\times M_{i2}\right) p \nonumber \\
&=&  V_{cd}V_{ud}  \left(M^2\right)_{22}p,  \label{eq:H_internal}
\end{eqnarray}
where $V_{cd}$ and $V_{ud}$ are the CKM matrix elements, and the flavor wave functions of the baryons are $p=u(ud-du)/\sqrt{2}$ and $\Lambda_c=c(ud-du)/\sqrt{2}$. 
$M$ is the matrix in terms of the pseudoscalar mesons, using the standard $\eta$-$\eta'$ mixing~\cite{Roca:2003uk}, 
\begin{eqnarray}
M&=&\left(\begin{array}{ccc}
              u\bar{u} & u\bar{d} & u\bar{s}\\
              d\bar{u} & d\bar{d} & d\bar{s} \\
              s\bar{u} & s\bar{d} & s\bar{s}
      \end{array}
\right) \nonumber \\
&\Rightarrow &
\left(\begin{array}{ccc}
              \frac{\pi^0}{\sqrt{2}}  + \frac{\eta}{\sqrt{3}}+\frac{\eta'}{\sqrt{6}}& \pi^+ & K^+\\
              \pi^-& -\frac{\pi^0}{\sqrt{2}} + \frac{\eta}{\sqrt{3}}+ \frac{\eta'}{\sqrt{6}}& K^0\\
               K^-& \bar{K}^0 & -\frac{\eta}{\sqrt{3}}+ \frac{2\eta'}{\sqrt{6}}
      \end{array}
\right) . \nonumber \\ \label{eq:Mmatrix}
\end{eqnarray}
Although the $a_0(980)$ could decay into the $\pi\eta'$ channel, we ignore the $\eta'$ component in this work as done in Refs.~\cite{Liang:2015qva,Xie:2014tma,Duan:2020vye,Wang:2021naf,Wang:2021nxz}, because the $\eta'$ has a large mass and does not play the role in the dynamical generation of the $a_0(980)$~\cite{Oller:1997chi}. As pointed out in Ref.~\cite{Oset:2016lyh}, the $\eta'$ component is omitted in the chiral Lagrangians because it is not a Goldstone Boson due to the $U_A(1)$ anomaly. 
Now, we have the possible components of the final states after the hadronization,
\begin{eqnarray}
H^{(a)}&=&  V_{cd}V_{ud}  \left(M^2\right)_{22}p \nonumber \\
&=&  V_{cd}V_{ud}  \left( \pi^+\pi^-+\frac{1}{2}\pi^0\pi^0-\frac{2}{\sqrt{6}}\pi^0\eta \right. \nonumber \\
&& \left. +\frac{1}{3}\eta\eta+K^0\bar{K}^0 \right)p \,,  \label{eq:H_a0}
\end{eqnarray}
where the coefficients account for the relative strengths of the different components. It should be pointed out that $H^{(a)}$ is the possible components of hadrons, instead of the effective Hamiltonian. 
Since the channels $\pi^+\pi^-$, $\pi^0\pi^0$, and $\eta\eta$ only couple to the system of isospin $I=0$ in the $S$-wave, and have no contribution for the $a_0(980)$ production in the $\Lambda_c$ decay, we eliminate the components of these three channels.  Equation~(\ref{eq:H_a0}) can be rewritten as,
\begin{eqnarray}
H^{(a)}&=&  V_{cd}V_{ud}  \left(M^2\right)_{22}p \nonumber \\
&=&  V_{cd}V_{ud}  \left( -\frac{2}{\sqrt{6}}\pi^0\eta +K^0\bar{K}^0 \right)p \,.  \label{eq:H_a}
\end{eqnarray}

For the mechanisms of Figs.~\ref{Fig:quarkthreebody}(c) and \ref{Fig:quarkthreebody}(d), where the $\bar{u}u$ pair is produced from the vacuum between $u$ from the $W^+$ decay and $ud$ quarks of the initial $\Lambda_c$, we have,
\begin{eqnarray}
\Lambda_c & \Rightarrow & V_{cd}V_{ud} d\bar{d}u\left(\bar{u}u\right)\frac{1}{\sqrt{2}}\left(ud-du\right) \nonumber \\
&=&  V_{cd}V_{ud} \left( M_{22}\times M_{11}\right)  \frac{1}{\sqrt{2}}u\left(ud-du\right)\nonumber \\ 
&=&  V_{cd}V_{ud} \left(-\frac{\pi^0}{\sqrt{2}}+\frac{\eta}{\sqrt{3}} \right)\left(\frac{\pi^0}{\sqrt{2}}+\frac{\eta}{\sqrt{3}} \right) p \nonumber \\
&=& V_{cd}V_{ud} \left(-\frac{1}{\sqrt{6}}\pi^0\eta + \frac{1}{\sqrt{6}}\pi^0\eta \right) p =0 ,  
\end{eqnarray}
for Fig.~\ref{Fig:quarkthreebody}(c), and 
\begin{eqnarray}
\Lambda_c & \Rightarrow & V_{cs}V_{us} s\bar{s}u\left(\bar{u}u\right)\frac{1}{\sqrt{2}}\left(ud-du\right) \nonumber \\
&=&  V_{cs}V_{us} \left( M_{33}\times M_{11}\right)  \frac{1}{\sqrt{2}}u\left(ud-du\right)\nonumber \\ 
&=&  V_{cs}V_{us} \left(-\frac{\eta}{\sqrt{3}} \right)\left(\frac{\pi^0}{\sqrt{2}}+\frac{\eta}{\sqrt{3}} \right) p \nonumber \\
&=& V_{cs}V_{us} \left(-\frac{1}{\sqrt{6}}\pi^0\eta  \right) p,  
\end{eqnarray} 
for Fig.~\ref{Fig:quarkthreebody}(d), respectively, where we have eliminated the components of $\pi^0\pi^0$ and $\eta\eta$ that only couples to the system of isospin $I=0$. Thus the mechanism of Fig.~\ref{Fig:quarkthreebody}(c) has no contribution to the $\pi^0\eta p$ production. For Fig.~\ref{Fig:quarkthreebody}(d), we have the components of the final states,
\begin{eqnarray}
H^{(d)}&=& V_{cs}V_{us} \left(-\frac{1}{\sqrt{6}}\pi^0\eta  \right) p. \label{eq:H_d}
\end{eqnarray}

Similar to the Figs.~\ref{Fig:quarkthreebody}(c) and \ref{Fig:quarkthreebody}(d), the $\bar{q}q$ pair can also be created between the $ud$ pair of the initial $\Lambda_c$, and hadronizes into the proton ($uud$) and $\pi^0$ ($\bar{d}d$). However, the $ud$ pair of the initial $\Lambda_c$ is the most attractive ``good" diquark, and this mechanism will destruct the strong diquark correlation, which is not favored, as discussed in Ref.~\cite{Miyahara:2015cja}. Thus, we neglect the contributions from those mechanisms because of the diquark correlation.

Although the final states $\pi^0\eta p$ can not be directly produced via the $W^+$ external emission of the $\Lambda_c$ decay,  the $\Lambda_c \to p K^+ K^-$ can proceed via the mechanism of the $W^+$ external emission as depicted in Fig.~\ref{Fig:quarkthreebody}(e), then undergoes the re-scattering $K^+K^- \to \pi^0\eta$, which finally gives rise to the final states $\pi^0\eta p$. For the process $\Lambda_c \to p K^+ K^-$, we have the possible component of the final states,
\begin{eqnarray}
H^{(e)}&=&  V_{cs}V_{us}  K^+K^- p, \label{eq:H_b}
\end{eqnarray}
where  $V_{cs}$ and $V_{us}$ are the CKM matrix elements related to the Cabibbo angle. We take $V_{cd}=V_{us}=-{\rm sin}\theta_c=-0.22534$ and $V_{ud}=V_{cs}={\rm sin}\theta_c=0.97427$~\cite{Wang:2020}.

\begin{figure}[tbhp]
\begin{center}
\includegraphics[height=3.5cm,width=6.0cm]{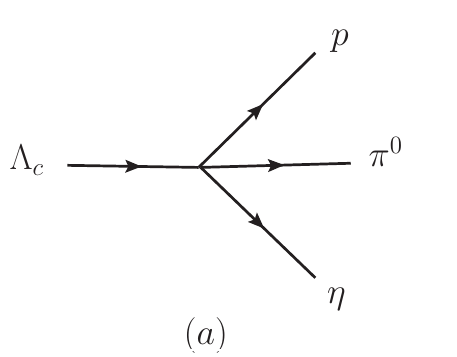}
\includegraphics[height=3.5cm,width=6.0cm]{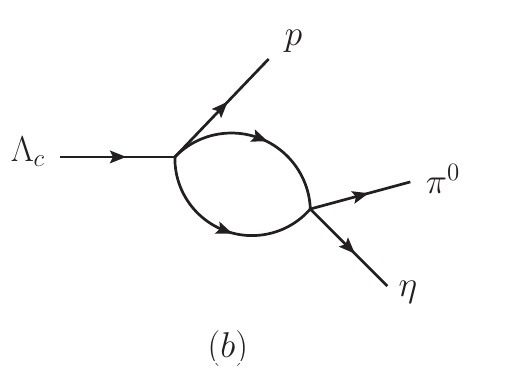}
\end{center}
\caption{The mechanisms of the decay $\Lambda_c \to  \pi^0\eta p$,
(a) tree diagram, (b) the $S$-wave  final state interactions.}
\label{Fig:feyman}
\end{figure}

For the process $\Lambda_c\to \pi^0\eta p$, in addition to the tree diagram of Fig.~\ref{Fig:feyman}(a), we also take into account  the $S$-wave pseudoscalar-pseudoscalar interaction, which will dynamically generate the scalar $a_0(980)$, as depicted in Fig.~\ref{Fig:feyman}(b). Taking into the contributions from Figs.~~\ref{Fig:quarkthreebody}(a), \ref{Fig:quarkthreebody}(d), and \ref{Fig:quarkthreebody}(e), the total amplitude for the $\Lambda_c\to \pi^0\eta p$ can be expressed as,
\begin{eqnarray}
\mathcal{M}&=& V_{p} V_{cd}V_{ud}\left(-\frac{3}{\sqrt{6}} - \frac{3}{\sqrt{6}}G_{\pi^0\eta}t_{\pi^0\eta\rightarrow \pi^0\eta} \right. \nonumber \\
&&\left.  +  G_{K^0\bar{K}^0}t_{K^0\bar{K}^0\rightarrow \pi^0\eta}  +  C\times  G_{K^+K^-}t_{K^+K^-\rightarrow \pi^0\eta}\right), \nonumber \\
\label{eq:amp}
\end{eqnarray}
where $V_{p}$ is the vertex of meson-meson production, which contains the weak amplitudes and the $q\bar{q}$ pair creation. 
 As discussed in Ref.~\cite{Geng:2018upx} where three-body charmed baryon decays are explicitly studied within the SU(3) flavor symmetry, the squared amplitude of the $\Lambda_c\to \pi^0\eta p$ without resonance effects is almost structureless, and the decay amplitude of the tree diagram is treated as a constant without energy dependence. Indeed, this hypothesis is widely used in the weak decays of the heavy hadrons~\cite{Miyahara:2015cja,Wang:2022xga}. Thus we take $V_p$ as a constant in here.

 We introduce  the color factor $C$ to account for the relative weight of the $W^+$ external emission mechanism with respect to the one of the $W^+$ internal emission mechanism, and the value of $C$ should be around 3 because we take the number of colors $N_c=3$~\cite{Wang:2020}\footnote{For the $W^+$ external emission of Fig.~\ref{Fig:quarkthreebody}(e), the $u\bar{s}$ quarks from the $W^+$ decay will hadronize into the color-singlet $K^+$, and the $u$ and $\bar{s}$ quarks could have three choices of the colors ($N_c=3$) because $W^+$ has no color.  However, for the $W^+$ internal emission of Fig.~\ref{Fig:quarkthreebody}(a), the $d$ quark from the $c$ decay has the same color as the $c$ quark, and the $\bar{d}$ and $u$ quarks from the $W^+$ decay should have the opposite and fixed colors, because the $u$ quark, together with the $ud$ of the initial $\Lambda_c$, has to match to the color singlet proton. It implies that the colors of all the final quarks are determined for the $W^+$ internal emission mechanism. 
Thus, the mechanism of the $W^+$ external emission (Fig.~\ref{Fig:quarkthreebody}(e)) is color-favored with respect to the ones of the $W^+$ internal emission (Figs.~\ref{Fig:quarkthreebody}(a)-(d)), and we introduce the color factor $C$ for the mechanism of the $W^+$ external emission.}. The diagonal matrix of the loop function $G_{l}$ is given by,
\begin{eqnarray}
G_{l}&=& i \int \frac{d^4 q}{(2\pi)^4}\frac{1}{(P-q)^2-m_1^2+i\epsilon}\frac{1}{q^2-m^2_2+i\epsilon}\nonumber \\ 
&=&  \int_0^{q_{\rm max}} \frac{|\vec{q}\,|^2d |\vec{q}\,|}{(2\pi)^2}\frac{\omega_1+\omega_2}{\omega_1
\omega_2} \frac{1}{s-(\omega_1+\omega_2)^2+i\epsilon}, \,
 \label{eq:loop}
\end{eqnarray}
where $s$ is the invariant mass squared of the meson-meson system, and $m_{1,2}$ are the meson masses. The meson energies $\omega_1=\sqrt{(\vec{q}\,)^2+m^2_1}$, $\omega_2=\sqrt{(\vec{q}\,)^2+m^2_2}$.
Since the $G_l$ is logarithmically divergent, we use the cut-off method to solve this singular integral, and take the cut-off parameter ${q}_{\rm max}=600$~MeV, as Ref.~\cite{Xie:2014tma}, in order to generate both the resonances $f_0(980)$ and $a_0(980)$ simultaneously.\footnote{It should be stressed that, in Ref.~\cite{Oller:1997chi}, the foot
$a_0(980)$, $f_0(980)$, and $f_0(500)$ 
resonances could be well generated from the unitary chiral approach with cut-off $q_{\rm max}=1.03$~GeV, as well as the $\pi\pi\to \pi\pi$, $\pi\pi\to K\bar{K}$ phase shifts and inelasticities in the $I=0$ scalar channel, and the $\pi^0\eta$ and $K\bar{K}$
 mass distribution in the $I=1$ channel.
 In Ref.~\cite{Liang:2014tia}, it is also shown that the cut-off $q_{\rm max}$ should be smaller than that of Ref.~\cite{Oller:1997chi}, when the $\eta\eta$ channel is explicitly taken into account in the $I = 0$ sector.
  Indeed, as shown by Fig.~4 of Ref.~\cite{Xie:2014tma}, the peak of $a_0(980)$ does not change too much with different values of $q_{\rm max}$ from 600~MeV to 800~MeV.}

The transition amplitude $t_{ij}$ in Eq.~(\ref{eq:amp}) is obtained by solving the Bethe-Salpeter equation in coupled channels,
\begin{equation}
T=[1-VG]^{-1}V,
\label{eq:bs}
\end{equation}
where three coupling channels $\pi^0\eta$, $K^+K^-$, and $K^0\bar{K}^0$ are included.
$V_{ij}$ represents the transition potential from $i$-channel to $j$-channel, as  Refs.~\cite{Xie:2014tma},
\begin{eqnarray}
&&  V_{K^+ K^- \to \pi^0 \eta}   = \frac{-
\sqrt{3}}{12f^2}(3s- \frac{8}{3}m^2_K-\frac{1}{3}m^2_{\pi}-m^2_{\eta} ), \\
&&V_{K^0\bar{K}^0 \to \pi^0 \eta} = - V_{K^+ K^- \to \pi^0 \eta} ,\\
&&V_{\pi^0 \eta \to \pi^0 \eta} = -\frac{1}{3f^2}m^2_{\pi}, \\
&&V_{K^+K^- \to K^+K^-} = -\frac{1}{2f^2} s , \\
&&V_{K^+K^- \to K^0 \bar{K}^0} = -\frac{1}{4f^2} s , \\
&&V_{K^0 \bar{K}^0 \to K^0 \bar{K}^0} =
-\frac{1}{2f^2} s ,
\end{eqnarray}
with the decay constant $f=f_\pi=93$~MeV. $m_\pi$ and $m_K$ are the isospin averaged masses of the pion and kaon, respectively.

With the full amplitude of Eq.~(\ref{eq:amp}), we can write the $\pi^{0}\eta$ invariant mass distribution for the process  $\Lambda_c\to \pi^0\eta p$,
\begin{equation}
\frac{d\Gamma}{dM_{\rm \pi^{0}\eta}}=\frac{1}{(2\pi)^3}\frac{p_p \tilde{p}_{\pi^{0}}}{4M^2_{\Lambda_c}}|\mathcal{M}|^2,
\label{eq:dw_threebody}
\end{equation}
where $p_p$ is the proton momentum in the $\Lambda_{c}$ rest frame,
\begin{equation}
p_p=\frac{\lambda^{1/2}(M^2_{\Lambda_c}, M^2_p, M^2_{\pi^0\eta}
)}{2M_{\Lambda_c}},
\end{equation}
$\tilde{p}_{\pi^{0}}$ is the $\pi^{0}$ momentum in the $\pi^0 \eta$ rest frame.
\begin{eqnarray}
\tilde{p}_{\pi^{0}} =
\frac{\lambda^{1/2}(M^2_{\pi^0\eta}, m^2_{\pi^0}, m^2_{\eta})}{2M_{\pi^0\eta}},
\end{eqnarray}
with K{\"a}llen function $\lambda(x,y,z)=x^2+y^2+z^2-2xy-2yz-2zx$. All the masses of the mesons and baryons involved in our calculations are taken from PDG~\cite{PDG2022}.

For the process $\Lambda_c\to \pi^0\eta p$, the rescattering of $\pi^0 p$ and $\eta p$ systems can give some contributions. Since there are many $N^*$ and $\Delta^*$ resonances, which could couple to the $\pi^0 p$ and $\eta p$~\cite{PDG2022}. We expect that those intermediate resonances will give a background contribution around 1~GeV in the $\pi^0\eta$ invariant mass distribution, and  not significantly change the position of $a_0(980)$. Thus, we neglect the rescattering of the  $\pi^0 p$ and $\eta p$  systems in this work, and will consider them when the experimental data are available in future.

\section{Results and Discussion}
\label{sec:results}

\begin{figure}[htpb]
\begin{center}
\includegraphics[scale=0.8]{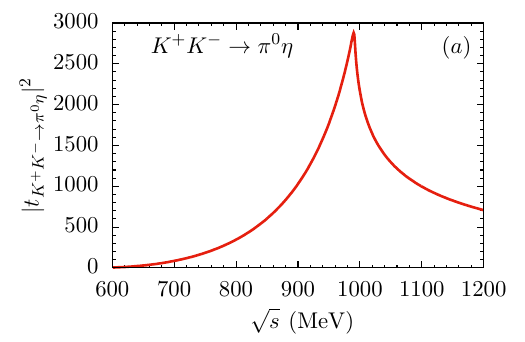}
\includegraphics[scale=0.8]{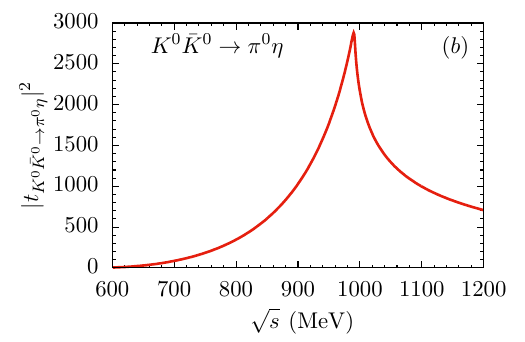}
\includegraphics[scale=0.8]{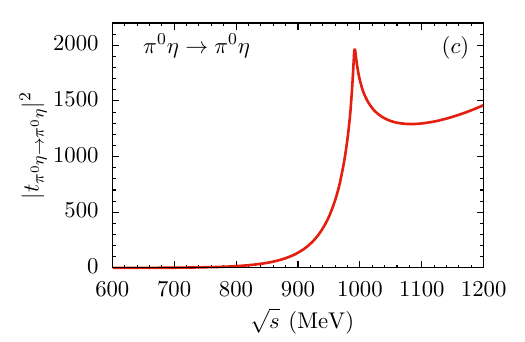}
\end{center}
\caption{The modulus squared of the amplitudes: (a) $|t_{K^{+}K^{-}\rightarrow \pi^{0}\eta}|^{2}$, (b) $|t_{K^{0}\bar{K}^{0}\rightarrow \pi^{0}\eta}|^{2}$, (c) $|t_{\pi^{0}\eta \rightarrow \pi^{0}\eta}|^{2}$.}
\label{Fig:t}
\end{figure}

\begin{figure}[htpb]
\begin{center}
\includegraphics[scale=0.8]{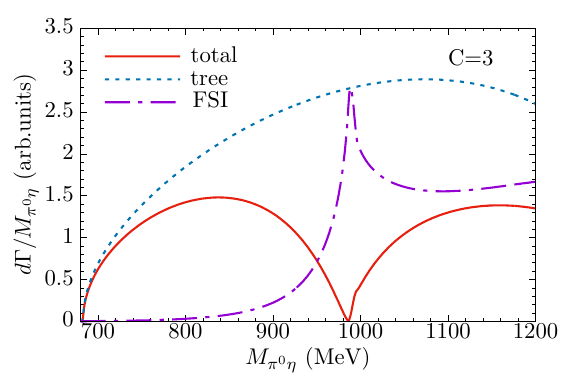}
\end{center}
\caption{The $M_{\pi^0\eta}$ invariant mass distribution of the process $\Lambda_c \to  \pi^0\eta p$. The purple dash-dotted curve shows the contribution from the $S$-wave pseudoscalar-pseudoscalar interaction as depicted in Fig.~\ref{Fig:feyman}(b), the blue dotted curve stands for the contribution from the tree level diagram as depicted in Fig.~\ref{Fig:feyman}(a), and the red solid curve corresponds to the total results given by Eq.~(\ref{eq:amp}).}
\label{Fig:dw}
\end{figure}

\begin{figure}[htpb]
\begin{center}
\includegraphics[scale=0.8]{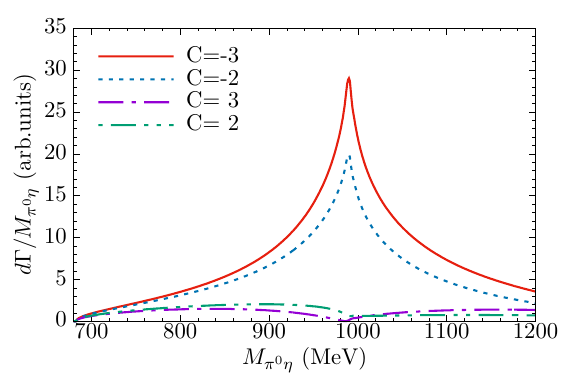}
\end{center}
\caption{The $M_{\pi^0\eta}$ invariant mass distribution for the process $\Lambda_c \to  \pi^0\eta p$ with different values of $C =$ 3, 2, -2, -3.}
\label{Fig:gamma}\end{figure}

In this section, we will present our results. In order to show the production of the  $a_0(980)$ in the $S$-wave $K^+K^-$, $K^0\bar{K}^0$, and $\pi^0\eta$ interactions, we respectively show the modulus squared $|t_{ij}|^2$ for the transitions $K^{+}K^{-}\rightarrow\pi^{0}\eta$, $K^{0}\bar{K}^{0}\rightarrow\pi^{0}\eta$, and $\pi^{0}\eta\rightarrow\pi^{0}\eta$ in Fig.~\ref{Fig:t}. From Fig.~\ref{Fig:t}, one can see a clear cusp structure around 980~MeV, which is associated to the $a_0(980)$. It should be pointed out that the cusp structure can be observed due to the threshold effect. However, due to the much more complicated situation, such as coupled channel with threshold near the resonance or the interference effect, this resonance may even shown up a cusp or a dip, as explained in Ref.~\cite{Guo:2019twa}. The cusp structure manifested by the  $a_0(980)$ is also supported by the experimental measurements~\cite{Debastiani:2017}.

In our model, we have two free parameters, $V_p$ and $C$. $V_p$ is a global factor and its value does not affect the line shape of the $\pi^0\eta$ invariant mass distribution. 
As mentioned above, the value of $C$ should be around 3. In the first step, we take $C=3$, and show the $\pi^0\eta$ invariant mass distribution up to an arbitrary normalization in Fig.~\ref{Fig:dw}. Instead of the cusp structure as shown in Fig.~\ref{Fig:t}, one can find a dip structure around 980~MeV in Fig.~\ref{Fig:dw}, which could be associated to the  $a_0(980)$. Although the  $a_0(980)$, as a dynamically generated state from the $S$-wave pseudoscalar-pseudoscalar, manifests as a cusp structure  in many processes, such as $\eta_c\to \eta \pi^+ \pi^-$~\cite{Debastiani:2017,Ablikim:2017}, it shows a dip structure in the process $\Lambda_c \to \pi^0\eta p$. 
As we know, hadron resonances are observed as narrow or broad peaks in the invariant mass distribution in many cases, a resonance may even show up as a dip, depending on the interference between the different contributions, as discussed in Ref.~\cite{Guo:2019twa}.  Indeed, this behavior is relatively common in hadron physics. For example, the $f_0(980)$ manifests itself as a clear peak in the $\pi^+\pi^-$ invariant mass distribution of the $J/\psi\to \phi K^+K^-$~\cite{Wu:2001vz,Augustin:1988ja}, and $B_s\to J/\psi \pi^+\pi^-$~\cite{Aaij:2011fx} reactions, but shows up as a dip in the $S$-wave $\pi\pi$ scattering amplitudes~\cite{Pelaez:2015qba}. The dip structures have also been found in experiments~\cite{Frabetti:2001ah,Aubert:2006jq}.

On the other hand, the $N_c$ scaling only indicates the relative strength of the absolute values, and the relative sign of $C$ is not fixed~\cite{Wang:2020}, thus we show the $\pi^0\eta$ invariant mass distribution with different values of $C = 3$, $2$, $-2$, $-3$ in Fig.~\ref{Fig:gamma}, where one can see a significant cusp structure around 980~MeV for the negative $C$, and a small dip structure for the positive $C$, which is due to the interference of the contribution from the final state interaction and the contribution from the tree diagram. It should be stressed that the  positive $C$ is supported by many studies, such as Ref.~\cite{Wang:2020}.  Thus, one dip or cusp structure in the $\pi^0\eta$ invariant mass distribution is expected to be observed in the more precision measurements for the process  $\Lambda_c\to \pi^0\eta p$.  Since our prediction for the $\pi^0\eta$ invariant mass distribution significantly depends on the sign of parameter $C$, the future precision measurements for this process could be used to constrain the parameter $C$, and deepen our understanding of the $N_c$ scaling in the weak decay of hadrons.

As we know, in the energy region around 1~GeV, there is only the state $a_0(980)$ that could decay into $\pi\eta$~\cite{PDG2022}. Thus if there is a peak/dip structure  observed in the $\pi\eta$ mass distribution around 1~GeV, it should be associated with the $a_0(980)$.

Since the process $\Lambda_c\to  \pi^0\eta p$ has not yet been measured experimentally up to our knowledge, we will predict the branching fraction of the process $\Lambda_c \to \pi^0\eta p$, which is important for the measurements of this process in the future.
The first step of the $\Lambda_c \to \pi^0\eta p$, weak process, is the same as the ones of the process $\Lambda_c \to p\pi^+\pi^-$ and $\Lambda_c \to p K^+K^-$~\cite{Wang:2020},  the normalization factor $V_p$, containing all dynamical factors of the weak process, is expected to be identical for these three processes. The branching fraction of the process $\Lambda_c\to \pi^0\eta p$ can be expressed as,
\begin{eqnarray}
\mathcal{B}(\Lambda_c \to \pi^0\eta p)&=& \frac{\int \left(\frac{d\Gamma}{dM_{\pi^0 \eta}}\right)dM_{\pi^0 \eta}}{\Gamma_{\Lambda_c}}.
\end{eqnarray}
It should be stressed that, the method of Ref.~\cite{Oller:1997chi} provides good amplitudes up to 1200~MeV~\cite{Molina:2019udw,Xie:2014tma,Debastiani:2017}, and one can not use the model for higher invariant masses. Thus, in this work, we take the integrating range of the $M_{\pi^0\eta}$ to be $(m_\pi+m_\eta)<M_{\pi^0\eta}<1200$~MeV.
 With the $V^2_p/\Gamma_{\Lambda_c}=0.2$~MeV$^{-1}$ from Ref.~\cite{Wang:2020}, we can roughly estimate the branching fraction $\mathcal{B}(\Lambda_c \to \pi^0\eta p)=(1.13\sim 1.26)\times 10^{-4} $ for  $2<C<3$, and $(5.96\sim 8.19)\times 10^{-4} $ for   $-3<C<-2$. The order of the magnitude $10^{-4}$ is expected to be accessible  in the BESIII and Belle II experiments. 



\section{Conclusions}
\label{sec:conc}
In this paper, the process $\Lambda_c\rightarrow \pi^0\eta p$ is investigated within the chiral unitary approach, where the contributions from the tree level diagram and the final state interactions of $K^{+}K^{-}$, $K^{0}\bar{K}^{0}$, and $\pi^{0}\eta$ are taken into account.

In our model, the interactions of the meson-meson pairs $K^+K^-$, $K^{0}\bar{K}^{0}$, and $\pi^0\eta$, preliminarily produced via the $\Lambda_c$ weak decay, result in the intermediate $a_0(980)$. We calculate the modulus square of the transition amplitudes for the $K^{+}K^{-}\rightarrow \pi^{0}\eta$, $K^{0}\bar{K}^{0}\rightarrow \pi^{0}\eta$, and $\pi^{0}\eta \rightarrow \pi^{0}\eta$ transitions, and find an obvious cusp near $980$~MeV, which corresponds to the $a_0(980)$ resonance. Up to an arbitrary normalization, we predict  the $\pi^0\eta$ invariant mass distribution, and find a dip structure for the positive values of $C$ and a cusp structure for the negative values of $C$.

As so far, no measurements of the process $\Lambda_c\rightarrow \pi^0\eta p$ were done, we roughly estimate the branching fraction $\mathcal{B}(\Lambda_c \to \pi^0\eta p)=(1.13\sim 1.26)\times 10^{-4} $ for the positive $C$, and $(5.96\sim 8.19)\times 10^{-4} $ for the negative $C$, which are expected to be accessible in the BESIII and Belle II experiments. 
Our rough estimation is smaller than the results of Ref.~\cite{Geng:2018upx} ($\mathcal{B}(\Lambda_c \to \pi^0\eta p)=(3.7\pm0.9)\times 10^{-3} $), and the reason should be that the different mechanisms are used in this work and Ref.~\cite{Geng:2018upx}. Thus the measurements of the branching fraction in future can be used to test the different results. On the other hand, the value of $V_p$, even containing the explicit dynamical factors, is expected to depend on the momentum smoothly, and does not significantly modify the lineshape of the $\pi^0\eta$ mass distribution.
Thus, we would like to call the attention of the experimentalists to measure this reaction, which should be useful to understand the nature of the $a_0(980)$, and the enhancement structure near the $K^+K^-$ threshold in the process $\Lambda_c\to p K^+K^-$ observed by the BESIII Collaboration~\cite{BESIII:2018mes}.

\begin{acknowledgments}
This work is supported by the National Natural Science Foundation of China under Grant No. 12192263, the Natural Science Foundation of Henan Province under Grant Nos. 222300420554, 232300421140, the Project of Youth Backbone Teachers of Colleges and Universities of Henan Province (2020GGJS017), and the Open Project of Guangxi Key Laboratory of Nuclear Physics and Nuclear Technology, No. NLK2021-08.
\end{acknowledgments}

  \end{document}